# Monitor backscatter factors for the Varian 21EX and TrueBeam linear accelerators: measurements and Monte Carlo modelling.

Sergei Zavgorodni[1,2], Eyad Alhakeem[2,1] and Reid Townson[2,1]

[1] *Department of Medical Physics, BC Cancer Agency – Vancouver Island Centre*
[2] *University of Victoria, Department of Physics & Astronomy*

Email: szavgorodni@bccancer.bc.ca

## 1. Abstract

**Purpose/Objective:** Linac backscattered radiation (BSR) into the monitor chamber affects the chamber's signal and has to be accounted for in radiotherapy dose calculations. In Monte Carlo (MC) calculations, the BSR can be modeled explicitly and accounted for in absolute dose. However, explicit modelling of the BSR becomes impossible if treatment head geometry is not available. In this study, monitor backscatter factors (MBSFs), defined as the ratio of the charge collected in the monitor chamber for a reference field to that of a given field, have been evaluated experimentally and incorporated into MC modelling of linacs with either known or unknown treatment head geometry.

**Materials and methods:** A telescopic technique similar to that by Kubo (1989) was used. However, instead of lead slits, a 1.8 mm diameter collimator and a small (2 mm diameter) detector positioned at extended SDD were used. This setup provided a field of view to the source of less than 3.1 mm and allowed for MBSF measurements of open fields from 1 x 1 to 40 x 40 $cm^2$. For the fields with both X and Y dimensions exceeding 15 cm, a diode detector was used. A pinpoint ionization chamber was used for smaller fields. MBSFs were also explicitly modeled in MC calculations using BEAMnrc and DOSXYZnrc codes for 6MV and 18MV beams of a Varian 21EX linac. A method for deriving the $D_{ch}^{forward}$ values that are used in MC absolute dose calculations was demonstrated. These values were derived from measured MBSFs for two 21EX and four TrueBeam energies.

**Results:** MBSFs were measured for 6MV and 18MV beams from Varian 21EX, and for 6MV, 10MV-FFF, 10MV, and 15MV beams from Varian TrueBeam linacs. For the open field sizes modeled in this study for the 21EX, the measured MBSFs agreed with MC calculated values within combined statistical (0.4%) and experimental (0.2%) uncertainties. Variation of MBSFs across field sizes was about a factor of two smaller for the TrueBeam compared to 21EX Varian linacs.

**Conclusions:** Measured MBSFs and the derived $D_{ch}^{forward}$ factors allow for the incorporation of the BSR effect into accurate radiotherapy dose calculations without explicit backscatter modelling.

## 2. Introduction

Backscattered radiation (BSR) from secondary collimators into the monitor chamber affects linac dose calibration and it has been investigated in the past both experimentally and using Monte Carlo (MC) modelling. Experimental measurement methods include using a backscatter filter technique (Luxton and Astrahan, 1988), the telescopic technique by Kubo (Kubo, 1989) and its variations, pulse counting and target charge measuring techniques. AAPM Task Group Report 74 (Zhu, Ahnesjo *et al.*, 2009) provides a concise but comprehensive summary of these studies.

To date, the telescopic technique has been recognized as the most reliable non-invasive method for measurement of the BSR (Yu, Sloboda *et al.*, 1996, Zhu, Ahnesjo *et al.*, 2009). However, it is cumbersome to use due to the heavy lead blocks required for collimation. Modifications to this technique that use smaller collimators have been tested (Duzenli, McClean *et al.*, 1993), but suffered from overestimation of the BSR due to inadvertent inclusion of room scatter into the measured signal (Yu, Sloboda *et al.*, 1996). Yu and Sanz (Yu, Sloboda *et al.*, 1996, Sanz, Alvarez *et al.*, 2007) improved this technique by evaluating and subtracting the scatter component of the signal, though Sanz *et al* used a small lead slug instead of collimators. Evaluation of the room scatter required doubling the number of measurements.

Understanding and evaluating the BSR impact on dose calculations has so far been largely of academic interest – in clinical dose calculation this factor has been incorporated and "hidden" within the collimator factor $S_c$ that is commonly measured during treatment machine commissioning. $S_c$ contains the monitor backscatter factor $S_b$ as well as the head scatter factor $S_h$ in the product $S_c = S_h \times S_b$, but neither $S_h$ or $S_b$ are used explicitly. Instead, the total scatter factor $S_{c,p}$ is also measured, and the phantom scatter factor (that is used in the calculations) is derived as their ratio: $S_p = S_{c,p} / S_c$.

In contrast, none of these empirically derived factors are required for use in Monte Carlo based dose calculations, as their effect is explicitly modelled during radiation transport through the linac head and into the phantom. However, the importance of $S_b$ factor measurements has risen dramatically with increased use of manufacturers' phase-spaces as well as simplified source models in place of full radiation transport through the linac head. Explicit modelling of the BSR is impossible or impractical in such situations and experimental $S_b$ values must be used instead. Measurement of the $S_b$ factor then becomes one of the key elements in absolute dose calculations.

The purpose of this study is to evaluate experimentally and incorporate into MC modelling the monitor backscatter factors (MBSFs), defined as the ratio of the charge collected in the beam monitor chamber for a reference field to that of a given field.

The novelty of the current paper is as follows: (1) improvement and simplification of the telescopic technique compared to its previous versions; (2) validation of the measurement technique by independent Monte Carlo modelling of $S_b$ factors for the Varian 21EX clinac (for which the geometry and composition was modeled from the manufacturer's specifications); (3) measurement of $S_b$ factors for Varian TrueBeam linear accelerator beams; (4) development of a simple technique for the derivation of forward fluence dose to the linac monitor chamber; (5) calculation of this forward fluence dose for two 21EX and four TrueBeam energies.

## 3. Materials and Methods

### *2.1 Monitor backscatter factors*

In the early publications by Kubo and Duzenli, the effect of the BSR was quantified simply as "relative charge reading" or "relative output". Later, Liu *et al* (Liu, Mackie *et al.*, 2000) introduced the factor $S_{cb}$, defined as "the change of photon output caused by the backscatter", which was subsequently simplified to the "monitor-backscatter factor" $S_b$ by Zhu *et al* (Zhu, Kang *et al.*, 2006, Zhu, Ahnesjo *et al.*, 2009). We will use the same terminology for consistency with the recent literature.

Zhu *et al* (2006) used the definition of $S_b$ in the BSR pulse measuring technique as the ratio of the number of electron pulses *NP(X,Y)* measured on the target for the field size *(X,Y)* to that for a reference field $(X_r,Y_r)$. AAPM task group report 74 (Zhu, Ahnesjo *et al.*, 2009) defines $S_b$ through a less intuitive ratio

$$S_b = \frac{\left(1 + b(A_{ref})\right)}{\left(1 + b(A)\right)}, \quad (1)$$

where $A$ and $A_{ref}$ are aperture settings for an arbitrary and reference field, respectively, $b=MU_b/MU_0$, and with $MU_b$ and $MU_0$ being backscatter and direct monitor chamber signals. Noticing that direct signal is independent of the field size, this equation simplifies to a much more intuitive definition as the ratio of the signal from the beam monitor chamber for a reference field to that of a given field:

$$S_b = \frac{\left(MU_0 + MU_b\right)_{ref}}{\left(MU_0 + MU_b\right)_{field}} = \frac{\left(MU_{total}\right)_{ref}}{\left(MU_{total}\right)_{field}} \quad (2)$$

This definition is inverted compared to the common definitions of other output factors that are defined relative to a reference signal. However, since monitor chamber signal is inverse to the detector reading collected in telescopic BSR measurements, this definition is consistent with the original “relative charge readings” by Kubo. It is also consistent with the relative measurements in our modified telescopic technique (which excludes room scatter from the signal). In the telescopic technique this factor is therefore directly measured as

$$S_b = \frac{\left(Detector\ reading\right)_{field}}{\left(Detector\ reading\right)_{ref}} \quad (3)$$

### *2.2 Experimental setup and measurements*

We used a telescopic technique similar to that by Kubo (Kubo, 1989). Instead of narrow slits, a set of decommissioned Brainlab circular stereotactic collimators was used. The height and outer diameter of each collimator was 11.4 and 6.7 cm, respectively. One of the collimators was modified in-house to provide an aperture size of 1.8 mm in diameter, and it was attached to the treatment head using a standard Brainlab stereotactic collimator holder. The distance from the bottom of this collimator to the source was 73.5 cm. Four other collimators were stacked on the couch at its lowest position, with the smallest (5 mm aperture) collimator being closest to the source. This produced a well-type structure to act as a detector housing and provide shielding from head scattered radiation and room scatter. A PTW pinpoint ionization chamber (effective volume 0.0125 $cm^3$ and effective diameter 2 mm) with a small PMMA build-up cap was positioned inside this well-type structure at a 150 cm source to detector distance (SDD), as shown in figure 1. The chamber was tightly fitted to the aperture of the lower collimator. This setup allows for accurate alignment of the apparatus in the beam (using the light field and lasers) and was found to provide high reproducibility of the measurements.

The in-field portion of the chamber cable was shielded by lead blocks to remove extra-chamber signal. Despite this, it was found that lead shielded cable current for large (over 15 x 15 $cm^2$) square fields still contributed up to 0.9% to the measured MBSFs for beams with energies of 10 MV and over. These large fields were therefore measured using a diode detector with a 2 mm diameter effective measurement volume. The diode detector is appropriate for MBSF measurements in our setup because only the signal from the narrow collimated beam is measured, with no spectral variations occurring over different field sizes.

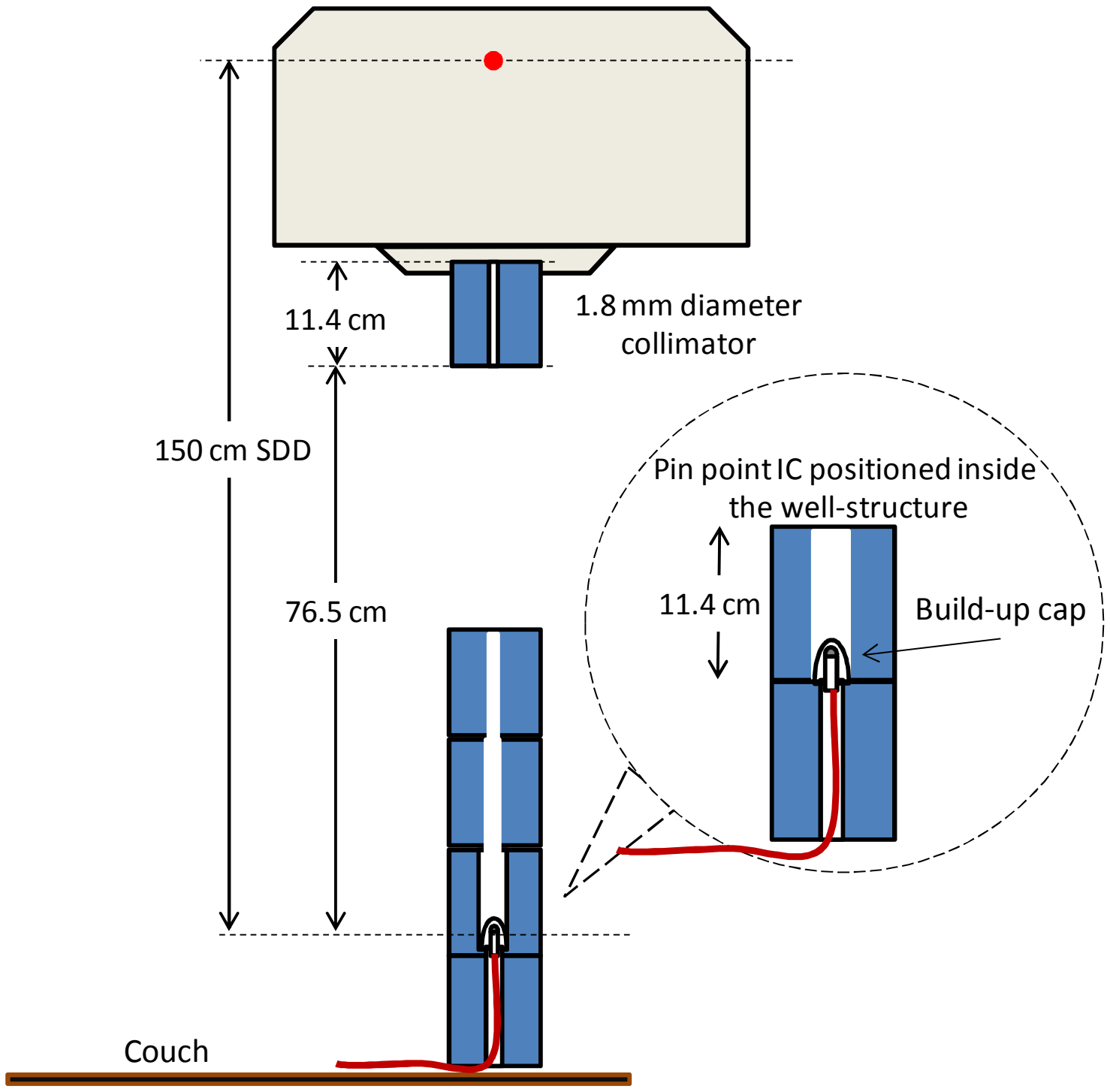

**Figure 1.** The experimental setup for measurement of $S_b$ factors. The 1.8 cm diameter collimator that was attached to the treatment head using a BrainLab SRS collimator holder is shown. Also shown are four collimators stacked on the couch top, providing a well-type structure for housing a PinPoint ionization chamber. The aperture of the top collimator was 5 mm. Inset: zoom-in of the IC with a small PMMA build-up cap, positioned inside the well-structure. Lead blocks that were used for chamber cable shielding are not shown. For large fields, a diode detector was used in place of the ionization chamber shown in the diagram.

This setup provided a field of view to the source of less than 3.1 mm, allowing for the measurement of $S_b$ factors for field sizes in the range of 1 x 1 to 40 x 40 cm$^2$.

In order to evaluate the level of the room scatter in the measured signal, a 2 cm diameter and 39 cm length steel rod was placed vertically on the top of detector housing collimator stack, blocking the primary signal. Measurements were taken for the highest available energy, 18 MV, for 10 x 10 and 40 x 40 cm$^2$ fields. Neither of these measurements detected any signal confirming negligible level of the room scatter contribution to the measured MBSFs.

One of the advantages of this setup was its high accuracy and reproducibility. The Brainlab stereotactic collimator holder has alignment screws to accurately centre the collimator in the beam. Once alignment was completed during the first measurement session, it was highly reproducible during subsequent sessions that were needed for measuring $S_b$ factors for different beams. The setup time for these sessions was on the order of 20 minutes.

The primary source of errors in these measurements was found to be signal drift due to instability of the machine output and pressure / temperature changes. Frequent (every 5 - 10 measurement points) reference measurements were taken to reduce effect of this drift and minimize measurement error.

Reproducibility of the setup and the measured data were verified by doing “spot check” measurements for all of the beams on different days (15MV and 18MV data sets have been re-measured completely). Additionally, these measurements were performed using a different detector (diode vs. PinPoint ionization chamber).

Experimental measurements were performed for 6MV and 18MV beams from the Varian 21EX, and for 6MV, 10MV, 10MV-FFF and 15MV TrueBeam linac beams. Three readings were taken per field size at 100 monitor units each. These readings, along with reproducibility measurements were used to evaluate uncertainties in the reported $S_b$ values.

An in-house sliding average filter was applied to the measured data to remove non-physical anomalies where $S_b$ factors decreased slightly with increasing field size. These corrections were very small, with majority of them being well within 0.1%. Only two points out of all the datasets (for six energies) were altered by 0.12%.

### *2.3 Incorporation of $S_b$ factors into Monte Carlo dose calculations*

In MC dose calculations the $S_b$ factor can be explicitly accounted for, provided the linac head geometry is known (Verhaegen, Symonds-Tayler *et al.*, 2000, Popescu, Shaw *et al.*, 2005). In fact, in our Vancouver Island MC (VIMC) system (Zavgorodni, Bush *et al.*, 2007, Bush, Townson *et al.*, 2008) we use the absolute dose calculation method (Popescu, Shaw *et al.*, 2005) that accounts for $S_b$ "on-the-fly". In this method, the dose is converted from MC dose (normalised per "incident electron on the target"), to absolute dose in units of Gy through the following equation:

$$D_{xyz,abs} = D_{xyz}\left(\frac{D_{ch}^{forward} + D_{ch}^{back}(ref)}{D_{ch}^{forward} + D_{ch}^{back}(field)}\right)\frac{D_{xyz,abs}^{cal}}{D_{xyz}^{cal}}, \qquad (4)$$

where

- $D_{xyz}$ is the normalized dose (in Gy per incident particle) scored during MC simulation in an arbitrary *xyz* voxel of a water or patient CT phantom.
- $D_{xyz,abs}$ is the absolute dose (in Gy) deposited in the voxel where $D_{xyz}$ was scored.
- $D_{ch}^{forward}$ is the normalized dose (in Gy per incident particle) scored during MC simulation in the monitor chamber of the virtual linac due to direct fluence.
- $D_{ch}^{back}$ is the normalized virtual linac monitor chamber dose (in Gy per incident particle) due to the BSR.
- $D_{xyz,abs}^{cal}$ is the absolute dose (in Gy) at a reference point in calibration geometry.
- $D_{xyz}^{cal}$ is the normalized dose (in Gy per incident particle) scored during MC simulation at a reference point in calibration geometry.

Taking notice of equation (2) and the fact that monitor chamber signal is proportional to deposited dose, equation (4) becomes

$$D_{xyz,abs} = D_{xyz} \cdot S_b \cdot \frac{D_{xyz,abs}^{cal}}{D_{xyz}^{cal}}, \qquad (5)$$

where

$$S_b = \left(\frac{D_{ch}^{forward} + D_{ch}^{back}(ref)}{D_{ch}^{forward} + D_{ch}^{back}(field)}\right), \qquad (6)$$

and it can be evaluated in MC calculations as described in Popescu *et al* (Popescu, Shaw *et al.*, 2005).

In situations where explicit radiation transport through the linac head geometry is not possible due to lack of geometry and material specifications, measured $S_b$ factors could be used through look-up tables. This is a satisfactory approach, but would involve empirical approximations for asymmetric fields where measured $S_b$ factors may not exist. Instead, for "virtual TrueBeam" calibration we propose a simple and more elegant approach that avoids these problems and maintains the integrity of absolute dose calibration. This is achieved by using equation (4) again but with the monitor chamber forward fluence dose being derived from measured $S_b$ factors and modeled $D_{ch}^{back}$:

$$D_{ch}^{forward} = \frac{D_{ch}^{back}(ref) - S_b \times D_{ch}^{back}}{S_b - 1} \tag{7}$$

In cases where the manufacturer has not provided the chamber geometry, any "virtual chamber" can be instead simulated using this approach. In this work, the calibration values for TrueBeam phase-space sources (supplied by the manufacturer) were determined by simulating BSR into a model of a 21EX chamber. With this model, the $D_{ch}^{back}$ values were calculated as described in Popescu *et al* (Popescu, Shaw et al., 2005). As discussed previously by Popescu et al., exact monitor chamber specifications are not required for absolute dose conversion, and therefore the use of the 21EX chamber model does not compromise the accuracy of the calibration method. This approach simply implies that, had this chamber been mounted in the TrueBeam treatment head, our derived values of $D_{ch}^{forward}$ and $D_{ch}^{back}$ would have been produced (by that chamber) for the measured set of $S_b$ factors. $D_{ch}^{forward}$ values have been obtained for four TrueBeam energies (6MV, 10MV-FFF, 10MV, and 15MV) available in our clinic.

It was found that $D_{ch}^{forward}$ values are quite sensitive to experimental errors that are inherent in measured $S_b$ factors, as well as statistical uncertainties in MC simulations of $D_{ch}^{back}$. In order to obtain the values of $D_{ch}^{forward}$ that provided the best fit to the measured $S_b$ values, the following procedure was used for each beam energy. $D_{ch}^{back}$ values were simulated for a variety of field sizes, and the measured $S_b$ factors were used to calculate $D_{ch}^{forward}$ for each field size via equation (7). In an ideal situation with no experimental or statistical errors, all of the $D_{ch}^{forward}$ values would have been exactly the same. However, due to uncertainties in $S_b$ and $D_{ch}^{back}$, the $D_{ch}^{forward}$ values had considerable spreads (from 18% to over 80% for different energies). It is worthwhile to note that due to the form of equation (6) and the fact that $D_{ch}^{forward}$ is about two orders of magnitude greater than $D_{ch}^{back}$, the $S_b$ factors derived from this equation were not sensitive to exact value of $D_{ch}^{forward}$. Using any value of $D_{ch}^{forward}$ from the calculated sets would have produced agreement with the measured $S_b$ factors within 0.6%. However, it was possible to considerably improve this agreement. For each of the beam energies, a set of $S_b^{calc}$ factors was derived for a range of field sizes from modeled $D_{ch}^{back}$ values and an "approximate" $D_{ch}^{forward}$ value using equation (6) in a simple calculation spreadsheet. The mean of the calculated set of $D_{ch}^{forward}$ values was a good first approximation for a given energy. The value of $D_{ch}^{forward}$ was then altered iteratively until the best fit of $S_b^{calc}$ factors to the measured TrueBeam $S_b$ factors was achieved. As fitting criteria, the root-mean-square deviation between the calculated and measured $S_b$ values was minimized with no differences $|(S_b^{calc} - S_b)|$ exceeding 0.1%. This procedure provided the "optimized" values of $D_{ch}^{forward}$, which, when used in MC calculations, produced $S_b$ factors in close agreement with experiment.

### *2.4 Monte Carlo modelling of the BSR for comparison with measurement*

MC modelling of the BSR was performed using the BEAMnrc code with our 6MV and 18MV beam models of the Varian 21EX clinac. These models have been thoroughly benchmarked and results reported previously (Zavgorodni, Locke *et al.*, 2005, Bush, Zavgorodni *et al.*, 2009, Bush, Gagné *et al.*, 2011, Zavgorodni, 2013). Sets of 7 and 10 square fields (ranging from 1 x 1 cm$^2$ to 40 x 40 cm$^2$) were used for 6 and 18 MV beams, respectively (see Figures 2 and 3 in Results section), and $S_b$ factors were calculated using the setup and equations described previously (Popescu *et al.*, 2005). MC modelling of $S_b$ factors from the TrueBeam was not possible due to lack of manufacturer's specifications on any components above the jaws. For the TrueBeam machines, the BSR was modeled using 21EX monitor chamber specifications. For both the TrueBeam and the 21EX, phase-space sources were scored just above the jaws.

## 4. Results

### *3.1 Estimated uncertainties of measured $S_b$ factors*

Inter-reading variations of $S_b$ measurements were within 0.1% for all measured energies and field sizes with the TrueBeam generally providing more stable readings than the 21EX. Error-propagated uncertainty estimation of $S_b$ factors for TrueBeam remained at 0.1% level whereas it increased to be in the range of 0.1%-0.2% for 21EX beams. This estimate was confirmed by reproducibility measurements. All spot-checked values of Truebeam $S_b$ factors were reproduced within 0.1%, and 21EX $S_b$ values were reproduced within 0.2%.

Larger differences of up to 0.5% were found during reproducibility measurements for fields with an X or Y jaw opening of 1 cm. This was attributed to slightly inaccurate collimator centering in the initial set of measurements that resulted in the jaws partly shielding the primary beam signal. Data reported in tables 1-6 reflect more accurate setup that we believe is free from this artefact and have the same uncertainty as the rest of the table.

### *3.2 $S_b$ factors measured for 6MV and 18MV Varian 21EX beams*

Tables 1-2 show measured $S_b$ factors for symmetric 6MV and 18MV fields from the Varian 21EX clinac. The factors for a 6MV beam increase from 0.996 for a 1 x 1 cm$^2$ field to 1.014 for a 40 x 40 cm$^2$ field. Corresponding values for an 18MV beam are 0.995 and 1.020. The presented data provide sufficient information to be used in dose calculations with analytical head source models or to derive and verify $D_{ch}^{forward}$ factors for MC models as discussed in the section 2.3.

**Table 1.** $S_b$ factors measured for a 6MV Varian 21EX and a range of field sizes. Estimated uncertainty of the data was 0.2%.

| 6MV | | X, cm | | | | | | | |
|---|---|---|---|---|---|---|---|---|---|
| | | 1 | 2 | 3 | 6 | 10 | 15 | 25 | 40 |
| | 1 | 0.996 | 0.996 | 0.996 | 0.996 | 0.996 | 0.996 | 0.996 | 0.996 |
| | 2 | 0.996 | 0.996 | 0.997 | 0.997 | 0.997 | 0.997 | 0.997 | 0.997 |
| | 3 | 0.997 | 0.997 | 0.997 | 0.997 | 0.997 | 0.997 | 0.998 | 0.998 |
| Y, cm | 6 | 0.998 | 0.998 | 0.998 | 0.999 | 0.999 | 0.999 | 0.999 | 0.999 |
| | 10 | 0.999 | 0.999 | 0.999 | 1.000 | 1.000 | 1.000 | 1.001 | 1.002 |
| | 15 | 1.000 | 1.000 | 1.001 | 1.001 | 1.002 | 1.002 | 1.003 | 1.004 |
| | 25 | 1.001 | 1.003 | 1.004 | 1.005 | 1.005 | 1.006 | 1.007 | 1.009 |
| | 40 | 1.005 | 1.006 | 1.007 | 1.008 | 1.009 | 1.010 | 1.011 | 1.014 |

**Table 2.** $S_b$ factors measured for an 18MV Varian 21EX and a range of field sizes. Estimated uncertainty of the data was 0.2%.

| 18MV | | X, cm | | | | | | | |
|---|---|---|---|---|---|---|---|---|---|
| | | 1 | 2 | 3 | 6 | 10 | 15 | 25 | 40 |
| | 1 | 0.995 | 0.995 | 0.995 | 0.995 | 0.995 | 0.995 | 0.996 | 0.996 |
| | 2 | 0.995 | 0.995 | 0.996 | 0.996 | 0.996 | 0.996 | 0.996 | 0.996 |
| | 3 | 0.996 | 0.996 | 0.996 | 0.996 | 0.996 | 0.997 | 0.997 | 0.997 |
| Y, cm | 6 | 0.997 | 0.997 | 0.998 | 0.998 | 0.998 | 0.998 | 0.999 | 1.000 |
| | 10 | 0.999 | 0.999 | 0.999 | 1.000 | 1.000 | 1.001 | 1.001 | 1.003 |
| | 15 | 1.001 | 1.002 | 1.002 | 1.002 | 1.003 | 1.004 | 1.005 | 1.007 |
| | 25 | 1.004 | 1.005 | 1.005 | 1.006 | 1.007 | 1.007 | 1.009 | 1.014 |
| | 40 | 1.009 | 1.011 | 1.012 | 1.012 | 1.013 | 1.015 | 1.016 | 1.020 |

**Table 3.** $S_b$ factors measured for a 6 MV Varian TrueBeam and a range of field sizes. Estimated uncertainty of the data was 0.1%.

| **6MV** | | **X, cm** | | | | | | | |
|---|---|---|---|---|---|---|---|---|---|
| | | 1 | 2 | 3 | 6 | 10 | 15 | 25 | 40 |
| **Y, cm** | 1 | 0.988 | 0.988 | 0.988 | 0.988 | 0.988 | 0.988 | 0.988 | 0.989 |
| | 2 | 0.998 | 0.999 | 0.999 | 0.999 | 0.999 | 0.999 | 0.999 | 1.000 |
| | 3 | 0.998 | 0.999 | 0.999 | 0.999 | 0.999 | 0.999 | 1.000 | 1.000 |
| | 6 | 0.999 | 0.999 | 0.999 | 0.999 | 1.000 | 1.000 | 1.000 | 1.000 |
| | 10 | 0.999 | 0.999 | 1.000 | 1.000 | 1.000 | 1.000 | 1.000 | 1.001 |
| | 15 | 0.999 | 1.000 | 1.000 | 1.000 | 1.000 | 1.001 | 1.001 | 1.001 |
| | 25 | 1.000 | 1.000 | 1.001 | 1.001 | 1.001 | 1.001 | 1.001 | 1.002 |
| | 40 | 1.001 | 1.001 | 1.001 | 1.002 | 1.002 | 1.002 | 1.002 | 1.003 |

**Table 4.** $S_b$ factors measured for a 10MV-FFF Varian TrueBeam and a range of field sizes. Estimated uncertainty of the data was 0.1%.

| **10MV-FFF** | | **X, cm** | | | | | | | |
|---|---|---|---|---|---|---|---|---|---|
| | | 1 | 2 | 3 | 6 | 10 | 15 | 25 | 40 |
| **Y, cm** | 1 | 0.993 | 0.993 | 0.993 | 0.993 | 0.993 | 0.994 | 0.994 | 0.994 |
| | 2 | 0.997 | 0.998 | 0.998 | 0.998 | 0.998 | 0.998 | 0.998 | 0.998 |
| | 3 | 0.998 | 0.998 | 0.998 | 0.998 | 0.998 | 0.998 | 0.999 | 0.999 |
| | 6 | 0.998 | 0.999 | 0.999 | 0.999 | 0.999 | 0.999 | 0.999 | 1.000 |
| | 10 | 0.999 | 0.999 | 0.999 | 1.000 | 1.000 | 1.000 | 1.001 | 1.001 |
| | 15 | 1.000 | 1.000 | 1.000 | 1.001 | 1.001 | 1.001 | 1.002 | 1.002 |
| | 25 | 1.001 | 1.001 | 1.002 | 1.002 | 1.002 | 1.003 | 1.003 | 1.004 |
| | 40 | 1.002 | 1.003 | 1.003 | 1.004 | 1.005 | 1.005 | 1.005 | 1.006 |

**Table 5.** $S_b$ factors measured for a 10MV Varian TrueBeam and a range of field sizes. Estimated uncertainty of the data was 0.1%.

| **10MV** | | **X, cm** | | | | | | | |
|---|---|---|---|---|---|---|---|---|---|
| | | 1 | 2 | 3 | 6 | 10 | 15 | 25 | 40 |
| **Y, cm** | 1 | 0.993 | 0.993 | 0.993 | 0.993 | 0.993 | 0.993 | 0.994 | 0.994 |
| | 2 | 0.998 | 0.998 | 0.998 | 0.998 | 0.998 | 0.998 | 0.999 | 0.999 |
| | 3 | 0.998 | 0.998 | 0.998 | 0.999 | 0.999 | 0.999 | 0.999 | 0.999 |
| | 6 | 0.998 | 0.999 | 0.999 | 0.999 | 0.999 | 0.999 | 1.000 | 1.000 |
| | 10 | 0.999 | 0.999 | 0.999 | 1.000 | 1.000 | 1.000 | 1.000 | 1.001 |
| | 15 | 0.999 | 1.000 | 1.000 | 1.000 | 1.001 | 1.001 | 1.001 | 1.002 |
| | 25 | 1.000 | 1.000 | 1.001 | 1.001 | 1.002 | 1.002 | 1.003 | 1.003 |
| | 40 | 1.000 | 1.001 | 1.001 | 1.002 | 1.003 | 1.004 | 1.004 | 1.005 |

**Table 6.** $S_b$ factors measured for a 15MV Varian TrueBeam and a range of field sizes. Estimated uncertainty of the data was 0.1%.

| **15MV** | | **X, cm** | | | | | | | |
|---|---|---|---|---|---|---|---|---|---|
| | | 1 | 2 | 3 | 6 | 10 | 15 | 25 | 40 |
| **Y, cm** | 1 | 0.996 | 0.996 | 0.996 | 0.996 | 0.996 | 0.996 | 0.996 | 0.996 |
| | 2 | 0.997 | 0.997 | 0.997 | 0.998 | 0.998 | 0.998 | 0.998 | 0.999 |
| | 3 | 0.997 | 0.998 | 0.998 | 0.998 | 0.999 | 0.999 | 0.999 | 0.999 |
| | 6 | 0.998 | 0.998 | 0.999 | 0.999 | 0.999 | 1.000 | 1.000 | 1.000 |
| | 10 | 0.999 | 0.999 | 0.999 | 1.000 | 1.000 | 1.000 | 1.001 | 1.001 |
| | 15 | 0.999 | 1.000 | 1.000 | 1.000 | 1.001 | 1.001 | 1.002 | 1.002 |
| | 25 | 1.000 | 1.001 | 1.001 | 1.002 | 1.002 | 1.003 | 1.003 | 1.004 |
| | 40 | 1.001 | 1.002 | 1.002 | 1.003 | 1.004 | 1.004 | 1.005 | 1.006 |

### 3.3 $S_b$ factors measured for 6MV, 10MV, 10MV-FFF and 15 MV Varian TrueBeam beams

$S_b$ factors that were measured for the four TrueBeam energies (6MV, 10MV-FFF, 10MV and 15MV) available in our clinic are shown in tables 3-6. These tables illustrate that the variation of $S_b$ factors with field size for the TrueBeam is considerably less than for the 21EX clinac. For the lowest of our energies, 6MV, the values ranged from 0.988 (for 1 x 1 $cm^2$) to 1.003 (for 40 x 40 $cm^2$). For this energy it is worth noting that for the fields with Y-jaw openings exceeding 2 cm $S_b$ factors could reasonably be assumed as unity. The measured $S_b$ factors for 10MV-FFF and 10MV were nearly identical for all fields with Y-jaw opening less than 25 cm. In the case where Y=25 cm, 10MV-FFF $S_b$ factors exceeded those for 10MV beam by less than ~0.1%. For Y=40 cm, 10MV-FFF $S_b$ factors were greater than those for 10MV beam by only ~0.2%. This agrees with the previously reported similarity of 6MV and 6MV-FFF $S_b$ factors (Titt, Vassiliev *et al.*, 2006, Zhu, Kang *et al.*, 2006). For the largest energy of 15MV, $S_b$ factors ranged from 0.996 to 1.006.

### 3.4 Forward fluence values derived for two 21EX and four TrueBeam beams

The values of $D_{ch}^{forward}$ previously obtained by our group via full MC simulation of the 21EX clinac head (Popescu, Shaw *et al.*, 2005), were $2.46 \times 10^{-15}$ and $2.24 \times 10^{-14}$ Gy/$e^-$ for 6MV and 18MV beams, respectively. The "optimised" values obtained from measured $S_b$ factors as described in section 2.3 were $3.4 \times 10^{-15}$ and $2.1 \times 10^{-14}$ Gy/e. Despite the large differences between $D_{ch}^{forward}$ values obtained from full MC simulation and "optimised" ones, the resulting $S_b$ values agreed to within 0.5%. Both sets of $D_{ch}^{forward}$ values produced $S_b$ factors in agreement with experiment, as seen in figures 2-3. However, as expected, measurement-based $D_{ch}^{forward}$ values resulted in better agreement with the experimental data.

The $D_{ch}^{forward}$ values were optimised to match the measured TrueBeam $S_b$ factors, as described in section 2.3. These values are shown in Table 7. The table also shows the number of fields that were used to derive the measurement-based $D_{ch}^{forward}$, and the largest absolute deviation of the $S_b^{optimised}$ factors (calculated using the derived $D_{ch}^{forward}$ in equation (6)) from the measured $S_b$ ($S_b^{measured}$) factors.

**Table 7.** The $D_{ch}^{forward}$ values obtained for the Varian TrueBeam clinac. Also shown are the number of fields that were used to derive the measurement based $D_{ch}^{forward}$, and the largest absolute value of the difference between the measured $S_b$ factors ($S_b^{measured}$) and the ones calculated from $D_{ch}^{forward}$ ($S_b^{optimised}$).

| | **TrueBeam energies** | | | |
|---|---|---|---|---|
| | **6MV** | **10MV-FFF** | **10MV** | **15MV** |
| $D_{ch}^{forward}$ (Gy/e) | $1.8 \times 10^{-14}$ | $4.78 \times 10^{-14}$ | $4.38 \times 10^{-14}$ | $5.32 \times 10^{-14}$ |
| Number of fields used to derive $D_{ch}^{forward}$ | 25 | 9 | 17 | 12 |
| $\max\lvert(S_b^{measured} - S_b^{optimized})\rvert$ | 0.05% | 0.02% | 0.08% | 0.07% |

As seen in the table 7, the values of $D_{ch}^{forward}$ reported in this paper produced variability of $S_b$ factors not exceeding +/-0.1% for the range of field sizes from 1 x 40 to 40 x 1 $cm^2$, which is within experimental error of $S_b$ measurement.

### 3.5 Monte Carlo calculated $S_b$ factors for the Varian 21EX clinac

Figures 2 and 3 provide comparison of the measured $S_b$ factors for square radiation fields produced by the Varian 21EX clinac against Monte Carlo calculated values. The "MC initial" label is attributed to data obtained from full MC modelling with all factors required for equation (6) obtained from MC simulation. For a 6MV beam, agreement was within 0.25% for all field sizes except the largest one (40 x 40 $cm^2$), where the difference was 0.54%. For 18MV, agreement was within 0.12% for all fields. Given that the estimated uncertainty of the MC calculated MBSFs was 0.4% and uncertainty of measured values was 0.2%, the agreement is within combined MC and experimental uncertainties.

The proposed technique for determining $D_{ch}^{forward}$ values from measured $S_b$ factors (equation 7) was tested and found to reduce the difference between MC and experiment to be within 0.05% for all 6MV and 0.1% for 18MV fields.

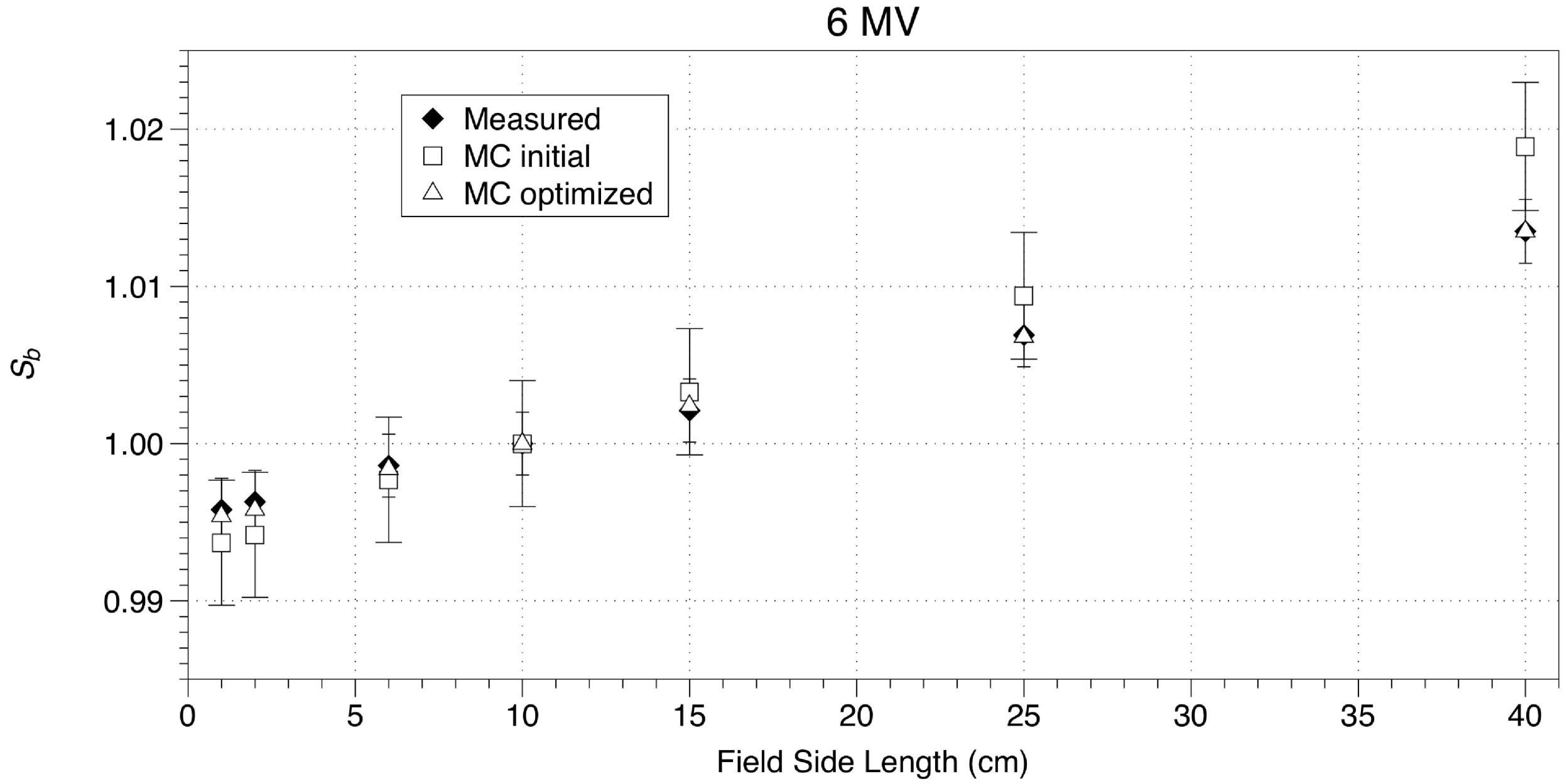

**Figure 2.** Comparison of measured and MC modelled $S_b$ factors for 6MV 21EX fields. Measured data are shown as diamonds. Full MC modelled (squares) $S_b$ factors were calculated using a $D_{ch}^{forward}$ value of 2.46x$10^{-15}$ Gy/$e^-$. The MC optimized (triangles) $S_b$ factors were derived using a $D_{ch}^{forward}$ value of 3.4x$10^{-15}$ Gy/e that was obtained through the fitting procedure described in section 2.3.

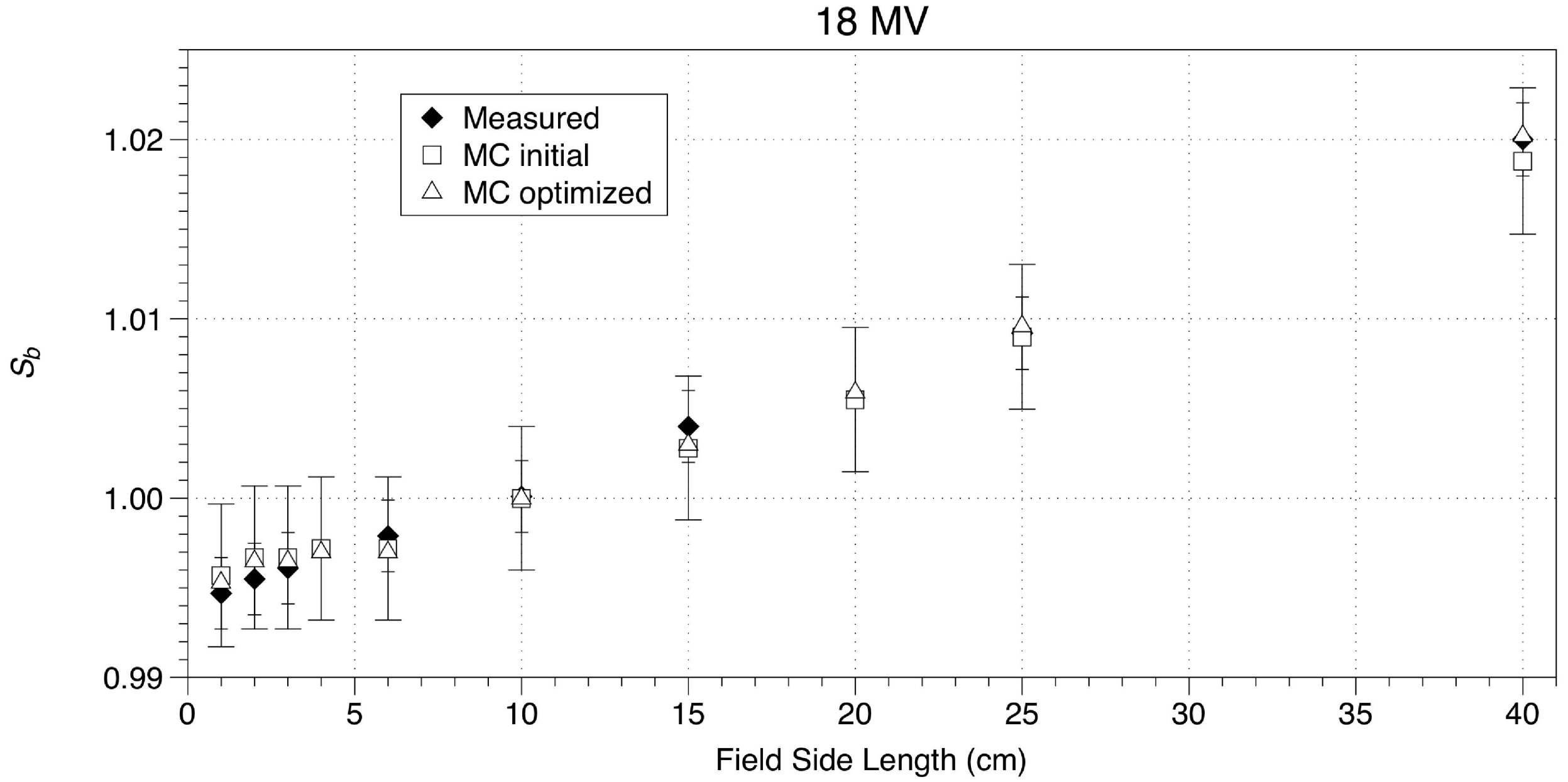

**Figure 3.** Comparison of measured and MC modelled $S_b$ factors for 18MV 21EX fields. Measured data are shown as diamonds. Full MC modelled (squares) $S_b$ factors were calculated using a $D_{ch}^{forward}$ value of 2.24x$10^{-14}$Gy/$e^-$. The MC optimized (triangles) $S_b$ factors were derived using a $D_{ch}^{forward}$ value of 2.1x$10^{-14}$ Gy/e that was obtained through the fitting procedure described in section 2.3.

## 5. Discussion

In this paper, an improvement of the telescopic technique for measuring the BSR to a monitor chamber was reported. Compared to the previously published telescopic techniques, our method avoided using bulky lead blocks (as in the method by Kubo), and allowed for the exclusion of room scatter from the measured signal without the need for repeated measurements as in Yu *et al*. and Sanz *et al*. (Yu, Sloboda *et al.*, 1996, Sanz, Alvarez *et al.*, 2007). Our method also enabled fast measurement setup that, after initial collimator alignment, only required about 20 minutes. The small (3.1 mm) field of view to the source, small detector size, and well-type design of the detector housing were all important components of this method that showed high reproducibility (0.1% for TrueBeam and 0.2% for 21EX linacs) of the measured factors.

The proposed experimental setup allowed MBSF measurements for open fields as small as 1 x 1 $cm^2$. Accurate alignment of the collimator to the source was found to be particularly important for the fields of this size as one of the upper jaws could partly shield primary signal. The technique could be further refined to reduce the measurable field size by extending the SDD, reducing detector size and reducing the collimator size. For larger fields accurate collimator centering becomes less critical because even if the collimator was not perfectly aligned to the center of the source, relative detector readings would still correctly reflect the BSR effect on radiation delivery.

The measured $S_b$ factors were compared against values obtained by MC modelling of Varian 21EX clinac beams. For the range of field sizes from 1 x 1 $cm^2$ to 40 x 40 $cm^2$ modeled in this study, measured MBSFs agreed with MC calculated values within 0.25%, with only one value difference (for 40 x 40 $cm^2$ field size, 6MV beam) being 0.54%, and this agreement was within combined statistical and experimental uncertainties (see figures 2 and 3).

The measured $S_b$ values for the 21EX clinac were also within 0.5% agreement with those reported previously (Titt, Vassiliev *et al.*, 2006, Zhu, Kang *et al.*, 2006), and within 0.5% with the values reported by Yu for similarly designed 2100CD (Yu, Sloboda *et al.*, 1996). Since the same technique was used for the TrueBeam $S_b$ factor measurements, we feel confident in the reported $S_b$ factors for this machine as well.

The MBSFs derived in this study removed up to 2.5% error for the 21EX and up to about 1% for the TrueBeam that would have been present in absolute dose calculations had these factors been ignored. Of particular note, these data demonstrate that the TrueBeam $S_b$ factors have reduced considerably compared to the 21EX clinac. In fact, for the 6MV beams with Y-jaw opening exceeding 1 cm, these factors could reasonably be assumed as unity. This conclusion is likely to also be valid for 6MV-FFF beams, as 6MV and 6MV-FFF beams were previously (Titt, Vassiliev *et al.*, 2006, Zhu, Kang *et al.*, 2006) shown to have a very similar magnitude of the BSR. Our data demonstrate the same similarity of $S_b$ factors between 10MV and 10MV-FFF beams. There is rather little $S_b$ variability (0.8%) for 10 MV and slightly more (1%) for 15MV beams. Due to the lack of manufacturer’s information on TrueBeam treatment head design we have no means to infer the cause of the reduced TrueBeam $S_b$ factors. Other manufacturers use designated anti-backscatter plates positioned between the jaws and monitor chamber for this purpose. Monitor chamber design and position along the beam axis would also affect measured MBSFs. It is possible that TrueBeam design includes some of these modifications.

In this paper we also demonstrate a simple technique for incorporating the measured $S_b$ factors into absolute dose calculations for the TrueBeam MC model (or any linac model with no available treatment head geometries). One advantage of our approach is that once the $D_{ch}^{forward}$ value is derived for a given beam energy, no further approximations are required for absolute dose calibrations even in the most complex dosimetric conditions. Our “virtual linac” calibration technique is not sensitive to the chamber geometry nor to the accuracy of $D_{ch}^{forward}$ values. This demonstrates the robustness of our MC dose calibration method that provides sub-percent absolute dose calculation accuracy even for relatively inaccurate values of the dose to the monitor chamber. In our 21EX model the change of a $D_{ch}^{forward}$ value by 38% resulted in less than 0.54% variation in $S_b$ factors.

Since the manufacturer has not specified the TrueBeam chamber design, we used chamber geometry from the 21EX clinac in the TrueBeam model. We are not claiming that the values of $D_{ch}^{forward}$ that we derived reflect the reality; they are simply a parameter of the MC model which provides accurate output calibration for a "virtual TrueBeam" MC model. Despite being "virtual", these values are likely to be useful to research groups that already have a model established for the 21EX clinac while being in the process of implementing TrueBeam modelling to their systems. In our case, the reported values of $D_{ch}^{forward}$ allowed for the reproduction of measured $S_b$ factors in MC TrueBeam modelling for 65 tested fields across four beam energies to within 0.1% (table 1).

The reported values of $S_b$ factors and $D_{ch}^{forward}$ should make the implementation of this method for accurate absolute dose calculations straightforward. Our MC absolute dose calculations for the TrueBeam model so far (data not shown here) have excellent (within 1%) agreement with experiment and Eclipse calculations in the regions where such good agreement is expected.

## 6. Conclusions

In this paper, the $S_b$ values measured for 6MV and 18MV beams from Varian 21EX and for four TrueBeam energies (6MV, 10MV-FFF, 10MV, 15MV) are reported. The $S_b$ factors were measured for a range of rectangular fields from 1 x 1 to 40 x 40 cm$^2$. This set of data should be useful for implementing $S_b$ factors into analytical as well as MC treatment head models. Our method of incorporating these factors into MC modelling allows for simple and natural derivation of a single required model parameter $D_{ch}^{forward}$ and avoids any empirical approximations while providing accurate absolute dose calibration under most complex dosimetric conditions.

**Acknowledgements**

We would like to acknowledge Stephen Gray for the precision manufacturing of the small collimator that was essential in obtaining the experimental results of this study. We would also like to acknowledge the contribution by University of Victoria undergraduate student Hilary Egglestone who helped to produce the MC results used in this study.